\documentclass[fleqn,usenatbib]{mnras}

\usepackage[T1]{fontenc}

\DeclareRobustCommand{\VAN}[3]{#2}
\let\VANthebibliography\thebibliography
\def\thebibliography{\DeclareRobustCommand{\VAN}[3]{##3}\VANthebibliography}

\usepackage{graphicx}	
\usepackage{amsmath}	
\usepackage{amssymb}	

\renewcommand{\vec}[1]{{\mathbfit #1}}

\newcommand{\pder}[2]{ \frac{\partial #1}{\partial #2} }
\newcommand{\pderN}[3]{ \frac{\partial^{#3} #1}{\partial #2^{#3}} }

\newcommand{\grad}{ {\bf \nabla } }
\newcommand{\curl}{ {\bf \nabla} \times}

\title{Propagating torsional Alfv\'en waves {in thermally active solar plasma}}

\author[S.A. Belov et al.]{
S.A. Belov$^{1,2}$\thanks{E-mail: mr\_beloff@mail.ru},
S. Vasheghani Farahani$^{3}$,
N.E. Molevich$^{1,2}$
\\
$^{1}$Department of Physics, Samara National Research University, Moscovskoe sh. 34, Samara, 443086, Russia\\
$^{2}$Department of Theoretical Physics, Lebedev Physical Institute, Novo-Sadovaya st. 221, Samara, 443011, Russia\\
$^{3}$Department of Physics, Tafresh University, Tafresh 39518 79611, Iran
}

\date{Accepted XXX. Received YYY; in original form ZZZ}

\pubyear{2021}

\begin{document}
\label{firstpage}
\pagerange{\pageref{firstpage}--\pageref{lastpage}}
\maketitle

\begin{abstract}
{The aim of the present study is to shed light on the effects connected with thermal misbalance due to non-equal cooling and heating rates induced by  density and temperature perturbations in solar active regions hosting either propagating torsional or shear Alfv\'en waves. A description for the nonlinear forces connected with Alfv\'en waves in non-ideal conditions is provided based on the second order thin flux tube approximation. This provides insight on the effects of Alfv\'en-induced motions on the boundary of thin magnetic structures in thermally active plasmas. The equations describing the process of generating longitudinal velocity perturbations together with density perturbations by nonlinear torsional Alfv\'en waves are obtained and solved analytically. It is shown that the phase shift (compared to the ideal case) and the amplitude of the induced longitudinal plasma motions against the period of mother Alfv\'en wave are greater for shear Alfv\'en waves compared to torsional Alfv\'en waves although following the same pattern. The difference in the influence of thermal misbalance on the induced velocity  perturbations is governed by the plasma-$\beta$ although its effect is stronger for shear waves. It is deduced that for a harmonic Alfv\'en driver the induced density perturbations are left uninfluenced by the thermal misbalance.} 

\end{abstract}




\section{Introduction}

{The solar atmosphere acts as a perfect laboratory for wave dynamic studies, especially for magnetohydrodynamic (MHD) waves which are ubiquitous in the solar atmosphere \citep{Nakariakov2020, 2021SSRv..217...76B}. In the present study, we focus on the propagation of the Alfv\'en wave since Alfv\'en waves have the potential to explain coronal heating and solar wind acceleration. For instance, there is evidence of upwardly propagating torsional Alfv\'en waves with the energy flux sufficient to heat the solar corona \citep{2009Sci...323.1582J,Srivastava2017} in addition to transferring energy released by solar flares. In particular, \citet{2020ApJ...891...99A} focused on solar flares and reported torsional oscillations at heights between the solar photosphere up to a maximum of $0.25$ solar radii with free energies around  $60\times 10^{23}$ J which provides insight on the amount of energy transfer due to torsional waves in massive flare events. Torsional Alfv\'en waves also feature in magnetic reconnection sites \citep{Kohutova2020,2020NatSR..1015603S}.}

{For MHD processes in the solar corona, it is important to account for the interplay of various thermodynamic processes maintaining the corona at reported temperatures. Moreover, for long-lived corona, continuous cooling and heating processes should be in equilibrium. In a ground breaking work, \cite{Field1965} showed that this thermal equilibrium is unstable under a wide range of conditions. As a result, three types of thermal instabilities can arise: isochoric, isobaric, and isentropic. In order to provoke such instabilities, the thermal balance between cooling and heating processes must experience perturbation. The fact of the matter is that when compressional plasma perturbations induced by MHD waves influence the thermal misbalance the concept of wave-induced thermal misbalance arises. The presence of wave-induced thermal misbalance makes the plasma media active for compessional waves to experience dispersion and amplification \citep{Molevich88, Nakariakov_2017}. For the solar corona, it was shown that characteristic times of the wave-induced thermal misbalance are as the order of the observed wave periods \citep{Zavershinskii2019}. The recent review about the wave-induced thermal misbalance in the solar corona is presented {by} \citet{Kolotkov2021}.}

{Among investigations of the influence of the wave-induced thermal misbalance on MHD waves, the major part is dedicated to the influence on slow waves in solar corona. \cite{DeMoortel2004} investigated the effect of optically thin cooling and constant heating on slow waves and compared it with the effect of thermal conduction. It was shown that the effect of thermal conduction was more pronounced for the considered parameters. \cite{Nakariakov_2017} considered a weakly nonlinear evolution of slow waves in flux tubes accounting for the thermal conduction and viscosity, and the thermal misbalance.{ The obtained results indicated that the thermal misbalance }could be important for {slow wave dynamics} in the solar corona, and could be used to explain certain discrepancies between theoretical results and observations. Moreover, \cite{Nakariakov_2017} proposed that slow waves could act for the diagnostics of the unspecified coronal heating function before \cite{Kolotkov_2020}   assumed that the stability of slow and entropy waves can constrain the coronal heating function. For standing slow magnetoacoustic waves in hot coronal loops, the thermal misbalance in the presence of thermal conduction was taken under consideration by \citet{2019A&A...628A.133K}, where the damping time associated with the thermal conduction was found to be an order of magnitude higher than the damping time associated with the thermal misbalance. Depending on specific heating and cooling functions, three regimes of slow wave evolution are proposed: acoustic over-stability, suppressed damping, and enhanced damping with damping rate near observational values. \citet{Zavershinskii2019} showed that the thermal misbalance can create quasi-periodic magnetoacoustic slow waves while its heating and cooling timescales dictate their characteristic periods. The estimated damping time of these slow wave trains in the solar corona is between $10$ and $100$ minutes \citep{2021A&A...646A.155D}. }

While thermal misbalance influences the dynamics of magnetoacoustic waves even in the linear regime, the Aflv\'en wave is only affected in the nonlinear regime which is due to the fact that Alfv\'en waves are incompressible in the linear regime. However they can generate compressional plasma motions in the nonlinear regime \citep{Hollweg1971,2011A&A...526A..80V}. \citet{Belov2021_Alf} showed that compressional plasma motions generated by shear Alfv\'en waves could be affected by the thermal misbalance. {This is reflected by the term which is exponentially dependent on the coordinate along the magnetic field appearing in the expression for the induced longitudinal motions. This term was called the exponential bulk flow.} In case of a sinusoidal shear Alfv\'en wave driver, the amplitude and the phase shift of the {induced longitudinal motions} become dependent on the driver frequency. Moreover, {plasma motions} generated by Alfv\'en waves can interact with the mother Alfv\'en wave, this interaction which is called self-interaction leads to Alfv\'en wave steepening \citep{Cohen1974, Verwichte1999,2012A&A...544A.127V}. The thermal misbalance can affect the self-interaction of shear Alfv\'en waves leading to changes in the steepening rate and the resulting amplitude of Alfv\'en waves \citep{Belov2020, 2021R&QE...63..694B}. The scale of the effects of self-interaction of Alfvén waves in different regions of the solar atmosphere in the presence of a thermal misbalance has been studied by \citet{Belov2020}.

In the present paper, a further step is taken to study the nonlinearly induced longitudinal motions due to Alfv\'en waves when thermal misbalance is associated with the solar plasma magnetic structure. \citet{Belov2021_Alf} studied the nonlinear effects connected with shear Alfv\'en waves in the presence of thermal misbalance, in particular, the deviations of phase shifts and  amplitudes of the induced longitudinal motions. Therefore, the aim here is to compare the phase shifts and amplitudes of the longitudinal motions induced by  the torsional Alfv\'en wave with the analogous {longitudinal motions} induced by the shear Alfv\'en wave, as well as, to check the efficiency of this generation due to various environmental effects like plasma-$\beta$, which is directly related to the magnetic field in the solar atmospheric conditions.

{In} the proceeding section II, the model, equilibrium conditions together with the governing thermal misbalance relations considered in the present study are stated. In section III,  the evolutionary equations governing the induced perturbations are obtained. The comparison between the effects of the nonlinear shear and torsional Alfv\'en wave drivers on the induced perturbations regarding the phase shifts and amplitudes are carried out in section IV. In section V, the conclusions are presented.

\section{Model and equilibrium conditions}
We use the following set of MHD equations \citep{Priest2014}: 
\begin{equation}
\frac{\partial{\rho}}{\partial{t}}+\nabla\cdot\left(\rho\vec{v}\right)=0\,,
\label{Cont_full}
\end{equation}
\begin{equation}
\rho\left(\frac{\partial\vec{v}}{\partial{t}}
+\left(\vec{v}\cdot\nabla\right)\vec{v}\right)
=-\grad\!P-\frac{1}{4\pi}\vec{B}\times\left(\curl\vec{B}\right),
\label{Motion_full}
\end{equation}
\begin{equation}
\frac{\partial{\vec{B}}}{\partial{t}}=\curl\left(\vec{v}\times\vec{B}\right),
\label{Induction_full}
\end{equation}
\begin{equation}
\nabla\cdot\vec{B}=0 \,,
\label{Div_full}
\end{equation}
\begin{multline}
C_{V}\rho\left(\frac{\partial{T}}{\partial{t}}
+\left(\vec{v}\cdot\nabla\right)T\right)-
\frac{k_{\mathrm{B}}T}{m}\left(\frac{\partial{\rho}}{\partial{t}}
+\left(\vec{v}\cdot\nabla\right)\rho\right)=\\=-\rho\,\!Q\!\left(\rho, T\right),
\label{Energy_full}
\end{multline}
\begin{equation}
P=\frac{k_\mathrm{B}}{m}\rho T\,,
\label{State_full}
\end{equation}
where $\rho$, $T$, and $P$ respectively represent the density, temperature, and pressure of the plasma, while $\vec{v}$ and $\vec{B}$ are vectors of the plasma velocity and magnetic field. The Boltzmann constant, the mean mass per volume, and the specific heat capacity at constant volume are respectively shown by $k_\mathrm{B}$, $m$, and $C_{V}$. The term  $Q\!\left(\rho, T\right)=L\!\left(\rho, T\right)-H\!\left(\rho, T\right)$ is the generalized heat-loss function \citep{Parker1953, Field1965}, {where $L\!\left(\rho, T\right)$ and $H\!\left(\rho, T\right)$ are the cooling and heating rates, respectively. Under steady-state conditions, we have $Q\!\left(\rho_0, T_0\right)=L\!\left(\rho_0,T_0\right)-H\!\left(\rho_0, T_0\right)=L_0-H_0=0$.}
 {We introduce the generalized heat-loss function in the form of $Q\!\left(\rho, T\right)$ due to the following reasons. First, the solar corona is composed of optically thin plasma. Thus, there is no need to solve the radiation transfer equation. As a result, cooling rate due to the radiation can be calculated in the form $L\!\left(\rho, T\right)$ (see Eq. \ref{loss_f} for the exact form) \citep[see for details][]{Dere1997,Delzanna2020chianti}. Second, since up-to-date no self-consistent model exists for coronal heating, we also assume it as a function of density and temperature $H\!\left(\rho, T\right)$. This statement is also applicable in the case of impulsive heating if the corresponding timescale is much shorter than the waves timescale. We believe that is a reasonable first step regarding the influence of plasma heating on wave dynamics. Dependencies of the cooling and heating rates on density and temperature enable the perturbations of thermodynamic variables to create thermal misbalance. The impact on the wave dynamics is of interest in the present study.} Equations (\ref{Cont_full})-(\ref{State_full}) are supplemented by the mechanical equilibrium condition on the tube boundary:
\begin{equation}
P+\frac{B^2}{8\pi}={P}^T_{ext},
\label{Boun_Cond_full}
\end{equation}
were ${P}^T_{ext}$ {is the sum of gas and magnetic pressures} of the external medium that acts on the tube boundary.

 By implementing the perturbation theory and investigating the axial symmetric motions ($\partial/\partial\phi=0$), we have kept terms up to the second order of smallness in the Taylor expansion as 
\begin{align}
&v_r\approx\alpha^2 v_{r2} = \tilde{v}_r,\; v_\phi\approx\alpha v_{\phi1}+\alpha^2 v_{\phi2}=\tilde{v}_\phi,\;v_z\approx\alpha^2 v_{z2} = \tilde{v}_z,\nonumber\\
&B_r\approx\alpha^2 B_{r2} = \tilde{B}_r,\;  B_\phi\approx\alpha B_{\phi1}+\alpha^2 B_{\phi2}=\tilde{B}_\phi,\nonumber\\
&B_z\approx B_0+\alpha^2 B_{z2}=B_0+\tilde{B}_z, \;  \rho\approx \rho_0+\alpha^2\rho_2=\rho_0 + \tilde{\rho}, \nonumber\\
&T\approx T_0+\alpha^2 T_{2}=T_0+\tilde{T},\; P\approx P_0+\alpha^2 P_{2}=P_0+\tilde{P},
\label{PT_Expansions}
\end{align}
where $\left|\alpha\right|\ll1$ is a small parameter, subscripts 1 and 2 denote the first and second order perturbations, respectively. In expansion (\ref{PT_Expansions}), we consider that there are no compressional perturbations of the first order, i.e. there is initially only a torsional Alfv\'en wave. With this expansion, Eqs. (\ref{Cont_full})--(\ref{State_full}) together with the mechanical equilibrium condition (\ref{Boun_Cond_full}) can be written as:
\begin{equation}
	\pder{\tilde{\rho}}{t}+\frac{1}{r}\rho_0\pder{r\tilde{v}_r}{r}+\rho_0\pder{\tilde{v}_z}{z}=0,
	\label{Cont_2nd_order}
\end{equation}
\begin{multline}
	\rho_0\pder{\tilde{v}_r}{t}-\frac{1}{r}\rho_0\tilde{v}_\phi^2=-\pder{\tilde{P}}{r}-\\-\frac{1}{4\pi}\left(\frac{1}{r}\tilde{B}_\phi\pder{r\tilde{B}_\phi}{r}-B_0\left(\pder{\tilde{B}_r}{z}-\pder{\tilde{B}_z}{r}\right)\right),
\label{Motion_r_2nd_order}
\end{multline}
\begin{equation}
	\rho_0\pder{\tilde{v}_\phi}{t}=\frac{B_0}{4\pi}\pder{\tilde{B}_\phi}{z},
	\label{Motion_phi_2nd_order}
\end{equation}
\begin{equation}
\rho_0\pder{\tilde{v}_z}{t}=-\pder{\tilde{P}}{z}-\frac{1}{4\pi}\tilde{B}_\phi\pder{\tilde{B}_\phi}{z},
\label{Motion_z_2nd_order}
\end{equation}
\begin{equation}
\pder{\tilde{B}_r}{t}-B_0\pder{\tilde{v}_r}{z}=0,
	\label{Induction_r_2nd_order}
\end{equation}
\begin{equation}
\pder{\tilde{B}_\phi}{t}-B_0\pder{\tilde{v}_\phi}{z}=0,
\label{Induction_phi_2nd_order}
\end{equation}
\begin{equation}
\pder{\tilde{B}_z}{t}+\frac{B_0}{r}\pder{r\tilde{v}_r}{r}=0,
\label{Induction_z_2nd_order}
\end{equation}
\begin{equation}
\frac{1}{r}\pder{r\tilde{B}_r}{r}+\pder{\tilde{B}_z}{z}=0,
\label{Div_2nd_order}
\end{equation}
\begin{equation}
C_{V}\rho_0\pder{\tilde{T}}{t}-
\frac{k_{\mathrm{B}}T_0}{m}\pder{\tilde{\rho}}{t}=-\rho_0\left(Q_{\mathrm{0\rho}}\tilde{\rho}+Q_{\mathrm{0T}}\tilde{T}\right),
\label{Energy_2nd_order}
\end{equation}
\begin{equation}
\tilde{P}=\frac{\mathrm{k_B}}{m}\left(\rho_0 \tilde{T}+\tilde{\rho} T_0\right),
\label{State_2nd_order}
\end{equation}
\begin{equation}
	\tilde{P}+\frac{B_0}{4\pi}\tilde{B}_z+\frac{1}{8\pi}\tilde{B}_\phi^2=\tilde{P}^T_{ext}.
	\label{Bound_cond_2nd_order}
\end{equation}
Note that in Eq. (\ref{Energy_2nd_order}), we have $Q_{0T}=\left.\partial Q/\partial T\right|_{\rho_0, T_0}$, $Q_{0\rho}=\left.\partial Q/\partial\rho\right|_{\rho_0, T_0}$. The second term on the RHS of Eq. (\ref{Motion_z_2nd_order}) is the nonlinear ponderomotive force which is responsible for the longitudinal induced plasma motions due to the Alfv\'en wave \citep{2011A&A...526A..80V,2012A&A...544A.127V,2017ApJ...844..148V-2}. Since, in the context of the present study, wave dynamics is considered in the long {wavelength} limit, we assume perturbations for which the condition $R\ll\lambda$ is fulfilled with $R$ and $\lambda$ respectively representing the tube radius and characteristic wave length for the density, velocity, and magnetic field perturbations  along the tube. This allows implementing the second order thin flux tube approximation \citep{Zhugzhda96}. Under this approximation, the two dimensional problem can be reduced to a one dimensional problem where the variations in the radial direction are out of the question due to the thin geometry of the flux tube. The Taylor expansions of the physical quantities in the radial direction are \citep{Zhugzhda96}
\begin{align}
&\tilde{v}_r\approx Vr,\; \tilde{v}_\phi\approx\Omega r,\; \tilde{v}_z\approx u,\;\nonumber\\
&\tilde{B}_r\approx b_r r,\; \tilde{B}_\phi\approx J r,\;  \tilde{B}_z\approx b_z + b_{z2}r^2,\;\nonumber\\
&\tilde{\rho}\approx\rho,\; \tilde{T}\approx T,\;  \tilde{P}\approx p + p_2 r^2.
\label{Taylor_series}
\end{align}
We should mention that from hereafter we imply the parameters introduced in the Taylor series represented by  Eq. (\ref{Taylor_series}), which must not be mistaken with the full expressions for the quantities expressed in Eqs. (\ref{Cont_full})--(\ref{State_full}). Using the Taylor expansion (\ref{Taylor_series}) for Eqs. (\ref{Cont_2nd_order})--(\ref{Bound_cond_2nd_order}) and following the same procedure as for \citep{Zhugzhda96,2011A&A...526A..80V}, we can obtain the final governing system as:
\begin{equation} 
\pder{\rho}{t}+2\rho_0 V +\rho_0\pder{u}{z}=0,
\label{Cont}
\end{equation}
\begin{multline}
p+\frac{B_{0}}{4\pi}b_z-\frac{A_0}{2\pi}\left[\rho_0\left(\pder{V}{t}-\Omega^2\right)+\frac{1}{4\pi}\left(J^2+\frac{1}{2}B_0\pderN{b_z}{z}{2}\right)\right]=\\=\tilde{P}^T_\mathrm{ext},
\label{motionV}
\end{multline}
\begin{equation}
\rho_0\pder{u}{t}=-\pder{p}{z}-\frac{A_0}{4\pi^2}J\pder{J}{z},
\label{motionU}
\end{equation}
\begin{equation}
\rho_0\pder{\Omega}{t}=\frac{B_0}{4\pi}\pder{J}{z},
\label{motionOmega}
\end{equation}
\begin{equation}
\pder{J}{t}-B_0\pder{\Omega}{z}=0\,,
\label{inductionJ}
\end{equation}
\begin{equation}
\pder{b_z}{t}+2VB_0=0\,,
\label{inductionBz}
\end{equation}
\begin{equation}
C_{V}\rho_0\pder{T}{t}-\frac{\mathrm{k_B} T_0}{m}\pder{\rho}{t}=-\rho_0\left(Q_{\mathrm{0\rho}}\rho+Q_{\mathrm{0T}}T\right),
\label{energy}
\end{equation}
\begin{equation}
p=\frac{\mathrm{k_B}}{m}\left(\rho_0 T+\rho T_0\right).
\label{State}
\end{equation}
Here, $A_0=\pi R_0^2$ is the unperturbed tube cross-section. Comparing Eqs. (\ref{Cont})--(\ref{State}) with the classical second order thin flux tube approximation, one may find an additional nonlinear term on the RHS of Eq. (\ref{motionU}) which is absent in the classical approach. To obtain this term, we implemented the expansion of Eq. (\ref{Taylor_series}) on Eq. (\ref{Motion_z_2nd_order}) and evaluated it on the tube boundary ($r=R_0$). Under this approach, the term $R_0^2\partial p_2/\partial z$ also comes into play. However, we neglect this term in comparison with the term $\partial p/\partial z$, since using the thin flux tube approach implies the condition $R\ll\lambda$ to be fulfilled. As a result, the system obtained can be used on tube boundary (or in small vicinity of the boundary) not in the full tube volume. The set of equations expressed by Eqs. (\ref{Cont})--(\ref{State}) enables studying a solar plasma cylindrical magnetic structure while experiencing the thermal misbalance. In case of ideal conditions, it is instructive to see \citep{2011A&A...526A..80V}.

Now that the governing set of equations are described (Eqs. (\ref{Cont})--(\ref{State})), the basis for discussing how the thermal misbalance affects the induced plasma motions due to torsional Afv\'en waves is provided. 

\section{Plasma motions induced by torsional waves}\label{s:PlasmaMotions}
Combine Eqs. (\ref{motionOmega}) and (\ref{inductionJ}) to obtain the wave equation for torsional Alfv\'en waves
\begin{align}
\hat{D}_A J=0,\;\hat{D}_A=\pderN{}{t}{2}-C_A^2\pderN{}{z}{2},\; C_A^2=\frac{B_0^2}{4\pi\rho_0}.
\label{AlfvWaveEq}
\end{align}
Here, $C_A^2$ is the square of the Alfv\'en speed. The form of Eq. (\ref{AlfvWaveEq}) shows that there is no influence on Alfv\'en waves from Alfv\'en induced motions up to the second order. The full exact solution of Eq. (\ref{AlfvWaveEq}) can be represented as $J=J_1\left(z-C_A t\right)+J_2
\left(z+C_A t\right)$, where $J_1$ and $J_2$ are arbitrary functions. In our work, we consider only waves propagating in the positive direction. Thus, we have $J=J\left(\xi\right)$, where $\xi=z-C_A t$. 

Other equations from the system (\ref{Cont})--(\ref{State}) can be combined to derive equations describing how torsional Alfv\'en waves induce compressional plasma motions in plasma with the thermal misbalance. Let us focus our attention on longitudinal velocity and density perturbations as it may have implications in the context of Alfv\'enic winds.
\subsection{Longitudinal velocity}\label{ss:LongVel}
It is instructive to write the evolutionary equation based on the combined nonlinear effects and thermal misbalance. Thus, the equation describing the process of generation of longitudinal plasma velocity perturbations by torsional Alfv\'en waves has the form as 
\begin{multline}
\pder{}{t}\left[\rho_0\hat{D}u + \frac{A_0}{2\pi}C_S^2\frac{\partial^2}{\partial z\partial t}\left(\frac{J^2}{4\pi}-\rho_0\Omega^2\right)+C_S^2\frac{\partial^2 \tilde{P}^T_\mathrm{ext}}{\partial z\partial t}\right.+\\+\left.\frac{A_0}{4\pi^2}\left(\left(C_A^2+C_S^2\right)+\frac{A_0}{4\pi}\hat{D}_A\right)\pder{}{t}\left(J\pder{J}{z}\right)\right]+\\+\frac{1}{\tau_{V}}\left[\rho_0\hat{D}_Qu + \frac{A_0}{2\pi}C_{SQ}^2\frac{\partial^2}{\partial z\partial t}\left(\frac{J^2}{4\pi}-\rho_0\Omega^2\right)+C_{SQ}^2\frac{\partial^2 \tilde{P}^T_\mathrm{ext}}{\partial z\partial t}\right.+\\+\left.\frac{A_0}{4\pi^2}\left(\left(C_A^2+C_{SQ}^2\right)+\frac{A_0}{4\pi}\hat{D}_A\right)\pder{}{t}\left(J\pder{J}{z}\right)\right]=0,
\label{VelocityEqFull}
\end{multline}
where, in the first square-bracket term, the notations are introduced as 
\begin{align}
&\hat{D}=\left(C_A^2+C_S^2\right)\hat{D}_T+\frac{A_0}{4\pi}\hat{D}_S\hat{D}_A,\nonumber\\
&\hat{D}_S=\pderN{}{t}{2}-C_S^2\pderN{}{z}{2},\;\hat{D}_{T}=\pderN{}{t}{2}-C_{T}^2\pderN{}{z}{2},\nonumber\\
&C_S^2=\frac{C_P}{C_V}\frac{k_B\!T_0}{m}, \;C_P=C_V+\frac{k_B}{m},\;C_T^2=\frac{C_A^2 C_S^2}{C_A^2 + C_S^2},\nonumber
\end{align}
and, in the second square-bracket term, where all terms are proportional to the derivatives of the generalized heat-loss function $Q\left(\rho,T\right)$, we have 
\begin{align}
&\hat{D}_Q=\left(C_A^2+C_{SQ}^2\right)\hat{D}_{TQ}+\frac{A_0}{4\pi}\hat{D}_{SQ}\hat{D}_A,\; \tau_V=\frac{C_{V}}{Q_{0T}},\nonumber\\
&\hat{D}_{SQ}=\pderN{}{t}{2}-C_{SQ}^2\pderN{}{z}{2},\;\hat{D}_{TQ}=\pderN{}{t}{2}-C_{TQ}^2\pderN{}{z}{2},\nonumber\\
&C_{SQ}^2=\frac{\left(Q_{0T}T_0-Q_{0\rho}\rho_0\right)}{Q_{0T}T_0}\frac{k_B\!T_0}{m},\; C_{TQ}^2=\frac{C_A^2 C_{SQ}^2}{C_A^2 + C_{SQ}^2}.\nonumber
\end{align}
{ Here, we have introduced sound and tube speeds modified by thermal misbalance, $C_{SQ}$ and $C_{TQ}$, along with the standard sound and tube speeds, $C_S$ and $C_T$,} for more details see, e.g., \citep{Zavershinskii2019} for the modified sound speed and \citep{Belov2021} for the modified tube speed. We note that $\tau_V$ is the characteristic time associated with the thermal misbalance, see \citep{Zavershinskii2019,Kolotkov_2020} for details.

In both square brackets of Eq. (\ref{VelocityEqFull}), the nonlinear forces connected with the torsional Alfv\'en wave are present as well as the force connected with the back-reaction of the external media (the term proportional to the $\tilde{P}^T_\mathrm{ext}$). These nonlinear forces are the magnetic tension force (term proportional to $J^2/4\pi$), the centrifugal force (term proportional to $\Omega^2$), and the ponderomotive force (term proportional to $J\partial J/\partial z$) \citep[see e.g.][]{2011A&A...526A..80V}. The interplay between these nonlinear forces determines the properties and efficiency of plasma motion generation by torsional Alfv\'en waves. In case of nonlinear shear Alfv\'en waves, only the ponderomotive force features in the equation analogues to Eq. (\ref{VelocityEqFull}), see \citep{Belov2021_Alf}. 

Eq. (\ref{VelocityEqFull}) looks complicated, but it can be greatly simplified under several assumptions. First of all, we assume the external medium of the magnetic tube to be vacuum in a sense that although the equilibrium total magnetic pressure of the external medium exists but it is not perturbed. This means that we have $\tilde{P}^T_\mathrm{ext}=0$. Second, as assumed, we consider only Alfv\'en waves propagating in the positive direction, thus, it follows from Eqs. (\ref{motionOmega}) and (\ref{inductionJ}) that the magnetic tension and centrifugal forces cancel each other ($J^2/4\pi=\rho_0\Omega^2$), see also \citep{2011A&A...526A..80V}. Moreover, it follows from $J=J\left(z-C_A t\right)$ that $\hat{D}_A J^2=0$. Thus, for torsional Alfv\'en waves propagating in one (say positive) direction, only one term connected to the ponderomotive force retains in both square brackets of Equation (\ref{VelocityEqFull}). This term is a function of $\xi=z - C_A t$. It allows us to implement our final assumption, that is $u=u\left(\xi\right)$, which makes $\hat{D}_A u =0$. Note that only the motions induced by the Alfv\'en is desired in the present study.

 The assumptions impose Eq. (\ref{VelocityEqFull}) to transform into
\begin{multline}
\pder{}{t}\left(\rho_0\left(C_A^2+C_S^2\right)\hat{D}_Tu+\frac{A_0}{8\pi^2}\left(C_A^2+C_S^2\right)\frac{\partial^2 J^2}{\partial t\partial z}\right)+\\
+\frac{1}{\tau_{V}}\left(\rho_0\left(C_A^2+C_{SQ}^2\right)\hat{D}_{TQ}u+\frac{A_0}{8\pi^2}\left(C_A^2+C_{SQ}^2\right)\frac{\partial^2 J^2}{\partial t\partial z}\right)=0.
\label{VelocityEqShort}
\end{multline}
Equation (\ref{VelocityEqShort}) describes how torsional Alfv\'en waves induce longitudinal plasma motions by the ponderomotive force in a plasma structure with the thermal misbalance. The exact solution of Eq. (\ref{VelocityEqShort}) in the form $u=u\left(\xi\right)$ for the perturbations induced by propagating torsional Alfv\'en waves would be 
\begin{align}
&u=C e^{\Psi \xi} + K_1 \bar{J}^2 + K_2 e^{\Psi \xi}\int e^{-\Psi\xi}\bar{J}^2 d\xi,\label{u_solution}\\
&\bar{J}^2=\frac{R_0^2 J^2}{B_0^2},\; \Psi = \frac{1}{C_A \tau_{V}}, \; K_1 = \frac{C_A^2+C_S^2}{2 C_A},\; K_2 = \frac{C_S^2-C_{SQ}^2}{2 C_A^2 \tau_{V}}.\nonumber
\end{align}
The solution expressed by Eq. (\ref{u_solution}) for torsional waves has the same form as the solution for shear waves \citep{Belov2021_Alf}:
\begin{align}
&v_z=C e^{\Psi \xi} + K_1 \left(\frac{B_x}{B_0}\right)^2 + K_2 e^{\Psi \xi}\int e^{-\Psi\xi}\left(\frac{B_x}{B_0}\right)^2 d\xi,\label{vz_solution}\\
&\Psi = \frac{C_A^2-C_{SQ}^2}{C_A \tau_{V}\left(C_A^2-C_S^2\right)}, \; K_1 = \frac{C_A^3}{2\left(C_A^2-C_S^2\right)},\; \nonumber\\
&K_2 = \frac{C_A^2\left(C_S^2-C_{SQ}^2\right)}{2\tau_{V}\left(C_A^2-C_S^2\right)^2}.\nonumber
\end{align}

{It could be noticed that the difference between the cases of torsional and shear Alfv\'en waves is in the values of the coefficients $\Psi$, $K_1$, and $K_2$. More specifically, there is no term $\left(C_A^2-C_S^2\right)$  in the denominator for the case of torsional waves. This is a consequence of the fact that longitudinal motions induced by torsional Alfv\'en waves are less affected by the value of plasma $\beta$ \citep{2011A&A...526A..80V}. Moreover, \citet{Scalisi_2021} showed that the longitudinal motions induced by torsional Alfv\'en waves do not depend on plasma $\beta$.} 

In  Eq. (\ref{u_solution}), $C$ is an arbitrary constant {that its value can be obtained} from the condition $u\rvert_{\xi=0}=0$. {The dimensionless current density is represented as $\bar{J}$}. As stated for the solution of the shear wave \citep{Belov2021_Alf}, each of the three terms on the RHS of Eq. (\ref{u_solution}) plays a specific role regarding the induced perturbations due to torsional Alfv\'en waves. The first term on the RHS of Eq. (\ref{u_solution}) provides information regarding the exponential bulk flow. The second terms has already featured itself for ideal conditions as it is independent of the thermal misbalance, while the third term on the RHS possesses information regarding the thermal misbalance which would be absent for ideal conditions. 

{Let us} focus on the influence of the thermal misbalance for the case of a sinusoidal torsional Alfv\'en wave-driver as
\begin{equation}
\bar{J}=\alpha \sin\left(k\xi\right)=\alpha\frac{e^{i k \xi}-e^{-i k \xi}}{2i},\label{sin_alf}
\end{equation}
where $k$ represents the wave-number of the Alfv\'en wave and $\alpha$ represents the relative amplitude. Substitution of Eq. (\ref{sin_alf}) in the exact solution (\ref{u_solution}) yields
\begin{align}
&u = C e^{\Psi \xi} + U_0 - A \cos\left(2k\xi+\phi_0\right),\label{u_sol_for_sinus}\\
&C=\frac{\alpha^2}{4}\frac{4 \omega^2 \tau_V^2\left(C_S^2-C_{SQ}^2\right)}{C_A \left(1+4 \omega^2\tau_V^2\right)}, \;
U_0=\frac{\alpha^2}{4}\frac{C_A^2+C_{SQ}^2}{C_A},\nonumber\\
&A=\sqrt{A_1^2+A_2^2},\; \phi_0 = \arctan\left(\frac{A_2}{A_1}\right),\nonumber\\
&A_1 = \frac{\alpha^2}{4}\frac{\left(C_A^2+C_{SQ}^2\right)+4\omega^2\tau_{V}^2\left(C_A^2+C_S^2\right)}{C_A\left(1+4\omega^2\tau_V^2\right)},\nonumber\\
&A_2= \frac{\alpha^2}{4}\frac{2\omega\tau_V\left(C_{SQ}^2-C_S^2\right)}{C_A\left(1+4\omega^2\tau_V^2\right)},\nonumber
\end{align}
where $\omega = C_A k$ is the frequency of the Alfv\'en wave.

 It can be noticed from {the expression for constant $C$ in} Eq. (\ref{u_sol_for_sinus}) that the direction of the induced flows is determined by the relation between the squares of the sound and modified sound speeds respectively represented by $C_S^2$ and $C_{SQ}^2$. {By taking a further look, it could be deduced that if $C_S^2>C_{SQ}^2$, then the flow moves in the same direction as the Alfv\'en wave. This condition corresponds to the condition of isentropic stability $\tau_V\left(C_S^2-C_{SQ}^2\right)>0$, if $\tau_V>0$ \citep{Molevich88,Zavershinskii2020}. Thus, in the isentropically stable plasma, the flow is co-directed with the Alfv\'en wave and oppositely-directed in the isentropically unstable plasma.} The same result was obtained for the case of shear waves. {However regarding the shear Alfv\'en wave,} direction of the bulk flow depends also on the plasma-$\beta$. For the case considered in the present study, there is no dependence on the plasma-$\beta$. 
 
{ Also, it can be noticed that the dependence on frequency arises in the velocity amplitude $A$ of the oscillating part of Eq. \ref{u_sol_for_sinus}. The high- and low-frequency limits of the amplitude are}
\begin{equation} \label{amplimits}
A = \left\{  \begin{array}{ll} A_\mathrm{hf} = \frac{\alpha^2}{4}\frac{C_A^2+C_S^2}{C_A} & \textrm{for ~~} \omega\left|\tau_V\right|\gg1  \textrm{,}\\ A_\mathrm{lf} =  \frac{\alpha^2}{4}\frac{C_A^2+C_{SQ}^2}{C_A} & \textrm{for ~~} \omega\left|\tau_V\right|\ll1\textrm{.} \end{array} \right.
\end{equation} 
{It can be seen from Eq. (\ref{amplimits}) that, when $C_S^2>C_{SQ}^2$ we have $A_\mathrm{hf}>A_\mathrm{lf}$. It means that there is a more efficient generation of longitudinal oscillations with shorter periods in the isentropically stable plasma. In the opposite case of isentropic instability, we have $A_\mathrm{hf}<A_\mathrm{lf}$, which corresponds to  more efficient generation of longitudinal oscillations with longer periods.}

{The final feature, which can be found in Eq.  (\ref{u_sol_for_sinus}), is appearance of the frequency-dependent velocity phase shift $\phi_0$. The presence of a such phase-shift leads to state that the maxima of the induced flow and mother Alfv\'en waves are no longer coincides. In the isentropically stable plasma, the maxima of induced longitudinal perturbation overtakes the Alfv\'en wave maxima while they fall behind in the isentropically unstable plasma.}

\subsection{Density}\label{ss:Density}
In this subsection, we are going to investigate the influence of thermal misbalance on the generation of density perturbations. Analogous to Eq. (\ref{VelocityEqFull}), we can derive an equation that governs the density perturbation from the set equations expressed by Eqs. (\ref{Cont})--(\ref{State}): 
\begin{multline}
\pderN{}{t}{2}\left[\hat{D}\rho-\frac{A_0}{2\pi}\pderN{}{t}{2}\left(\frac{J^2}{4\pi}-\rho_0\Omega^2\right)-\pderN{\tilde{P}^T_\mathrm{ext}}{t}{2}\right.-\\
-\left.\frac{A_0}{4\pi^2}\left(C_A^2+\frac{A_0}{4\pi}\hat{D}_A\right)\pder{}{z}\left(J\pder{J}{z}\right)\right]+\\
+\frac{1}{\tau_{V}}\pder{}{t}\left[\hat{D}_Q\rho-\frac{A_0}{2\pi}\pderN{}{t}{2}\left(\frac{J^2}{4\pi}-\rho_0\Omega^2\right)-\pderN{\tilde{P}^T_\mathrm{ext}}{t}{2}\right.-\\
-\left.\frac{A_0}{4\pi^2}\left(C_A^2+\frac{A_0}{4\pi}\hat{D}_A\right)\pder{}{z}\left(J\pder{J}{z}\right)\right]=0.
\label{DensityEqFull}
\end{multline}
Equation (\ref{DensityEqFull}) has the same nonlinear forces as of Eq. (\ref{VelocityEqFull}). Following the same procedure as for the velocity, Eq. (\ref{DensityEqFull}) could be simplified to
\begin{multline}
\pderN{}{t}{2}\left(\left(C_A^2+C_S^2\right)\hat{D}_T\rho-\frac{A_0}{8\pi^2}C_A^2 \pderN{J^2}{z}{2}\right)+\\
+\frac{1}{\tau_V}\pder{}{t}\left(\left(C_A^2+C_{SQ}^2\right)\hat{D}_{TQ}\rho-\frac{A_0}{8\pi^2}C_A^2 \pderN{J^2}{z}{2}\right)=0.
\label{DensityEqShort}
\end{multline}
{Following the same logic as in the previous subsection for longitudinal plasma velocity, we can look for the} solution of Eq. (\ref{DensityEqShort}) in the form $\rho=\rho\left(\xi\right)$. Thus, the exact solution for the density perturbations induced by propagating torsional Alfv\'en waves would be 
\begin{equation}
	\rho = \frac{\rho_0}{2}\bar{J}^2+Ke^{\Psi\xi}.\label{densitySol}
\end{equation}
The solution expressed by Eq. (\ref{densitySol}) consists of two terms. The first term coincides with the solution for ideal plasma \citep{2011A&A...526A..80V} while the second term is connected with the mass flow introduced by the thermal misbalance. There is no additional influence of thermal misbalance in comparison to ideal conditions other than this effect. We note that $K$ is an arbitrary constant which is determined from the condition $\rho\rvert_{\xi=0}=0$, where if the Alfv\'en driver $J\rvert_{\xi=0}$ is equal to zero, we would have $K=0$, which means that there would be no mass flow. To state clearer, in case of a sinusoidal driver, we have
\begin{equation}
	\rho=\frac{\alpha^2}{4}\rho_0\left(1-\cos\left(2k\xi\right)\right),
\end{equation}
which shows that for the case of a propagating Alfv\'en wave driver, the induced mass flow is not affected by the thermal misbalance.

\section{Comparison between shear and torsional Alfv\'en drivers}\label{s:Comparison}

\begin{figure}
	\includegraphics[width= \columnwidth]{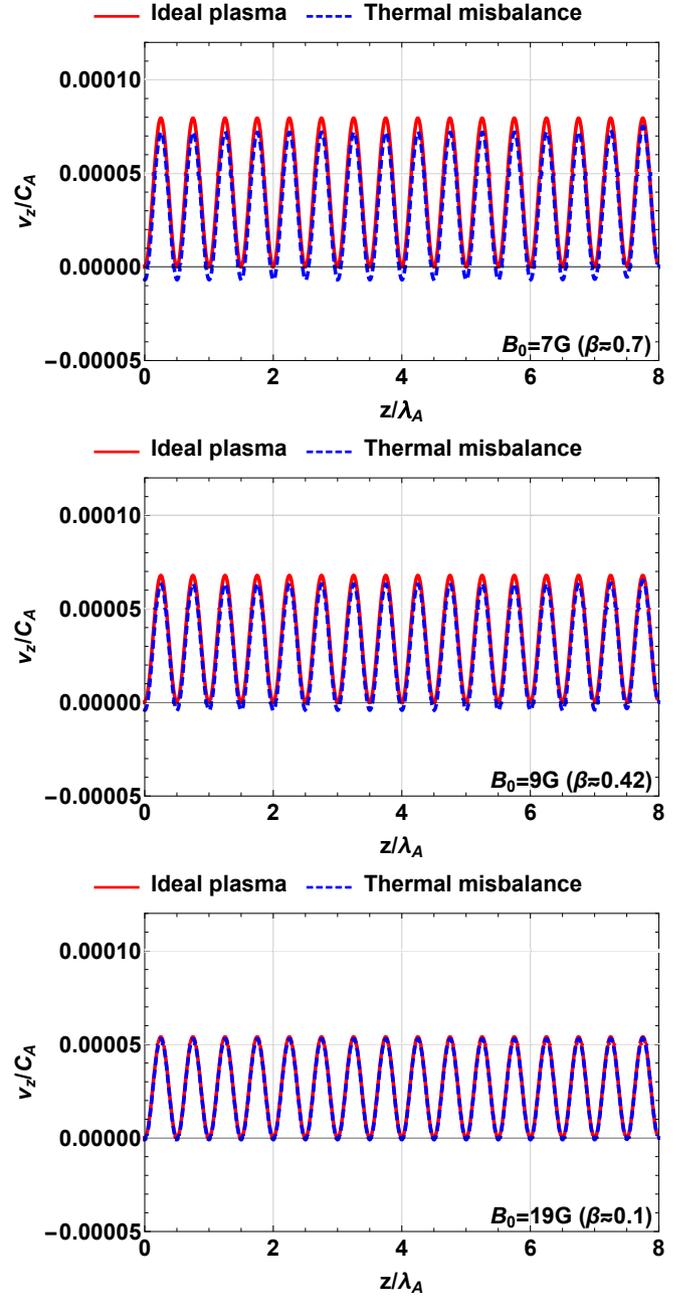}
	\caption{Relative amplitude of longitudinal plasma motion induced by torsional Alfv\'en waves in plasma with (red curve) and without (dashed-blue curve) thermal misbalance  for $T_0=1\,\mathrm{MK}$, $n_e=5\times10^9\,\mathrm{cm}^{-3}$, and $B_0=7\,\mathrm{G}, 9\,\mathrm{G}$, $19\,\mathrm{G}$. See also Fig. 3 of \citet{Belov2021_Alf}.}
	\label{fig:uComparison}
\end{figure}

\begin{figure*}
	\includegraphics[width= 17cm]{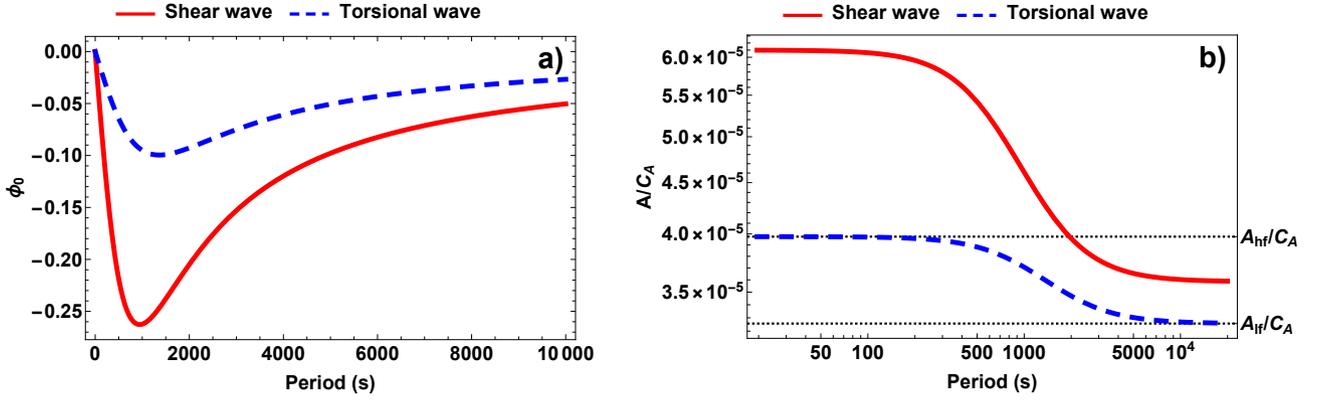}
	\caption{The effects of the Alfv{\'e}n wave period on two parameters, the phase shift and the amplitude. The red curves correspond to the shear wave driver while the dashed-blue curves correspond to the torsional wave driver. \textbf{Panel a)} The phase shift of the oscillating part with respect to ideal plasma conditions regarding longitudinal motions induced by the torsional and shear Alfv{\'e}n waves, see also Fig. 2 of \citet{Belov2021_Alf}. \textbf{Panel b)} The relative amplitude of the oscillating part with respect to ideal plasma conditions regarding longitudinal motions induced by the torsional and shear Alfv{\'e}n waves, see also Fig. 1 of \citet{Belov2021_Alf}. Note that we have considered $T_0=1\,\mathrm{MK}$, $n_e=5\times10^9\,\mathrm{cm}^{-3}$, and $B_0=7\,\mathrm{G}$ ($\beta\approx0.7$) }
	\label{fig:freqDep}
\end{figure*}

\begin{figure}
	\includegraphics[width= \columnwidth]{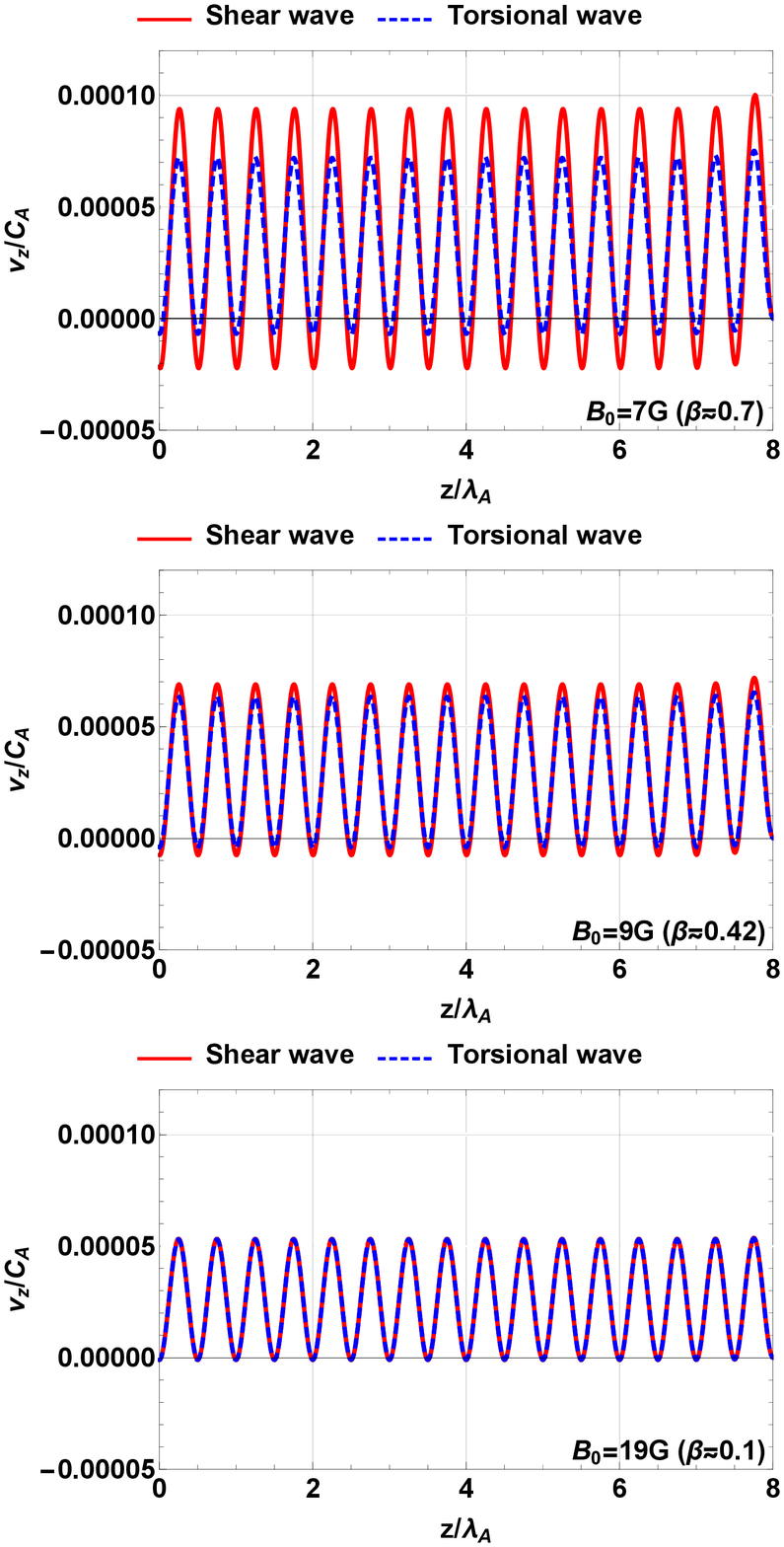}
	\caption{Relative amplitude of longitudinal plasma motion induced by shear (red curve) and torsional (dashed-blue curve) Alfv\'en drivers  for $T_0=1\,\mathrm{MK}$, $n_e=5\times10^9\,\mathrm{cm}^{-3}$, and $B_0=7\,\mathrm{G}, 9\,\mathrm{G}$, $19\,\mathrm{G}$. See also Fig. 3 of \citet{Belov2021_Alf}.}
	\label{fig:vzComparison}
\end{figure}

The mass flow generated by the torsional Alfv\'en wave remains the same as for the case of ideal plasma. Owing to this deduction, in this section, we restrict our attention only on the longitudinal plasma motions generated by torsional Alfv\'en waves and compare with their corresponding motions generated by shear Alfv\'en waves for plasma parameters associated with solar coronal conditions.

Before performing our comparison, we specify the heat-loss function as $Q\!\left(\rho, T\right)$. In the solar corona, radiation runs away without interacting with the plasma. This is why we only consider losses due to optically thin plasma radiation expressed as 
\begin{equation}
\label{loss_f}
L\!\left(\rho,T\right) = \frac{\rho}{4 m^2}\,\Lambda\!\left(T\right)\,,
\end{equation}
where $m =0.6\cdot 1.67\cdot10^{-24}\,g$ is the mean particle mass and $\Lambda\!\left(T\right)$ is the radiative-loss function determined from the CHIANTI atomic database v. 10.0 \citep{Dere1997,Delzanna2020chianti}. The heating function $H\!\left(\rho, T\right)$ can be locally modeled as
\begin{equation}
\label{heat_f}
H\!\left(\rho, T\right)=h\rho^aT^b\,,
\end{equation}
where $h$, $a$, and $b$ are given constants. The constant $h$ is determined from the steady-state condition $Q\!\left(\rho_0,T_0\right)=0$: $h=L\!\left(\rho_0,T_0\right)/\rho_0^a T_0^b$. The power-law indices $a$ and $b$ could be associated with some specific heating mechanism. {\cite{Kolotkov_2020} showed that for a plasma without thermal conduction there is a region of $-2\lesssim a\lesssim 2.5$ and $-5\lesssim b\lesssim -3$ where both thermal stability and acoustic stability exist in the coronal plasma. This means that criteria of isochoric, isobaric, and isentropic thermal stability are satisfied for $a$ and $b$ inside this region. In this parameter region, all solutions will be qualitatively the same, the difference will be in the current numerical values of heat-loss function derivatives $Q_{0T}$ and $Q_{0\rho}$, which implies the different values of the induced longitudinal plasma motions. But, once again, the general picture will be qualitatively the same. For example, the exponential bulk flow will be positive for all $a$ and $b$ in this region, but for isentropically unstable plasma, it will be negative. For illustrative purposes, we follow \cite{Kolotkov_2020} and also choose the point $a=1/2$, $b=-7/2$ inside this parameter region. We should mention that the calculations which will be provided in the proceeding are valid for the chosen region of parameters. However, the variation of the heating mechanism will lead to the change the calculated values.} 

{To highlight the influence of thermal misbalance on the induced plasma motions, we plot in Figure \ref{fig:uComparison} the longitudinal plasma motion induced by torsional Alfv\'en waves in plasma with and without thermal misbalance for Alfv\'en wave period $P_A=300$\,seconds and different plasma $\beta$ values. Also, we assume the plasma with temperature $T_0=1\,\mathrm{MK}$, {electron number density $n_e=5\times10^9\,\mathrm{cm}^{-3}$ which can correspond to the active region fan loop.} It should be mentioned that we start from the relatively higher values of $\beta\approx0.7$ because, according to \citep{Gary2001, Bourdin2017}, we can expect $\beta\sim1$ at relatively low coronal heights: {at $140\,\mathrm{Mm}$ for \citep{Gary2001} and at $50\,\mathrm{Mm}$ for \citep{Bourdin2017}}. We limit ourselves by only plotting the longitudinal velocity, because as we have shown, there should be no thermal misbalance influence on density perturbations for sinusoidal Alfv\'en driver, and moreover, in the present paper, we do not consider any influence from induced plasma motions on mother Alfv\'en waves. As can be seen from Figure \ref{fig:uComparison}, the main difference between these two cases is the appearance of negative values of the induced velocity perturbation. Also, there is small, but observable, "floating" of peaks in the case of thermal misbalance. These peaks are "floating up" due to the exponential bulk flow appeared because of the action of thermal misbalance. The same effect was investigated for the case of shear mother Alfv\'en wave \citep{Belov2021_Alf}. For the plasma $\beta$ decrease, the difference between plasma with and without thermal misbalance becomes less noticeable.}

As obtained in subsection \ref{ss:LongVel}, the oscillating part of the induced longitudinal plasma motion exhibits a phase shift with respect to the case of ideal plasma (see $\phi_0$ in Eq. (\ref{u_sol_for_sinus})). This phase shift is dependent on the frequency of the Alfv\'en wave driver. Plasma motions generated by shear Alfv\'en waves also experience the frequency dependent phase shift. Figure \ref{fig:freqDep}a compares this dependency for the case of shear and torsional Alfv\'en wave drivers. It can be readily noticed that for the case of torsional wave driver, the phase shift is much smaller, but the dependence is qualitatively the same as for the case of shear driver. It is instructive to compare the efficiency of the oscillating motion generated by torsional and shear Alfv\'en waves. Thus, we examine value of the amplitude of the oscillating part $A$ from Eq. (\ref{u_sol_for_sinus}) and compare it with the corresponding value for the case of shear wave driver, see Eq. (13) of \citet{Belov2021_Alf}. Figure \ref{fig:freqDep} demonstrates this comparison where, as well as for the phase shift, the amplitude of the plasma motion induced is less for the case of the torsional Alfv\'en driver. This  means that torsional Alfv\'en waves are less efficient for the generation of oscillating { longitudinal} plasma motions in comparison to shear waves.

Figure \ref{fig:vzComparison} compares the full profile of the induced plasma motion obtained from Eq. (\ref{u_sol_for_sinus}) with the profile for the case of shear driver for different values of the plasma-$\beta$ when the Alfv\'en wave period is set at $P_A=300$\,seconds. For the case where $\beta\approx0.7$, as was expected from the results, the plasma motion induced by torsional waves experiences a less phase-shift while observing a smaller amplitude than the motion induced by shear waves. Moreover, the exponential bulk flow is almost negligible. The difference between the relative values of the longitudinal plasma motions induced by torsional and shear waves { is negligible for $\beta\lesssim 0.1$}. In other words, the divergence between the induced amplitudes by the shear and torsional drivers are more pronounced for higher plasma-$\beta$ conditions. Thus, the efficiency of the thermal misbalance in inducing higher amplitude perturbations increases with the plasma-$\beta$.  Moreover, the thermal misbalance although affects the { longitudinal} motions induced by torsional Alfv\'en waves, but its influence is less than for the case of shear wave Alfv\'en wave drivers. {This is because of the geometrical difference between torsional and shear Alfv\'en waves drivers. Due to the axial symmetry, torsional waves produce less influence on plasma. For example, \citet{2011A&A...526A..80V} showed that for propagating torsional Alfv\'en waves, the efficiency of the nonlinear generation of compressible perturbations does not grow with the plasma-$\beta$ as for the case of shear waves. This result determines, why the influence of thermal misbalance is weaker for torsional Alfv\'en waves: there is no influence of plasma-$\beta$, and all this influence is due to appearance of the sound speed modified by the effect of thermal misbalance.}

\section{Conclusions}

In this work, the generation of longitudinal plasma motions by torsional Alfv\'en waves in thermally active plasma has been considered. For this purpose, a mathematical model describing nonlinear Alfv\'en waves and Alfv\'en-induced motions on the boundary of thin magnetic flux tube has been derived. In this approach, equations describing the process of generating longitudinal velocity perturbations together with density perturbations by nonlinear torsional Alfv\'en waves have been highlighted by obtaining and writing the corresponding evolutionary equation together with its solution.  

It has been demonstrated that for the case of torsional Alfv\'en waves the influence of thermal misbalance on the induced longitudinal plasma motion has the same features as for the case of shear Alfv\'en waves. An important feature in this regard is the presence of an exponential bulk flow besides the appearance of frequency-dependent velocity phase shifts. Another interesting aspect is that the velocity amplitude of the oscillating part of the induced motions is frequency-dependent, see also \citep{Belov2021_Alf}. Comparison with the case of the shear wave-driver for coronal conditions has revealed the fact that the efficiency of the thermal misbalance is less pronounced for torsional Alfv\'en waves. In addition, the efficiency of the thermal misbalance is more pronounced on the induced longitudinal motions for higher plasma-$\beta$ conditions although it features stronger for nonlinear shear Alfv\'en waves. 

Nonlinear torsional Alfv\'en waves induce density perturbations in ideal conditions \citep{2011A&A...526A..80V}. In the case of thermal misbalance, there appears an additional mass flow. However, for a wide class of wave-drivers, where Alfv\'en perturbations equal zero at the wave front, there is no mass flow and thus, { the thermal misbalance does not significantly affect the density perturbations induced by Alfv\'en waves}.

The intriguing result of this work is the difference in influence of the thermal misbalance  on plasma velocity and density perturbations. {The perturbation of density can even experience no influence due to the thermal misbalance. However, the  obtained difference between induced velocity and density perturbations} may be the result of simplicity of the model considered, i.e. due to not accounting for higher terms in radial expansion which could be crucial for the dynamics description. To provide a precise answer in this regard, we are motivated to consider more exact {2.5D models as in \citep{Scalisi_2021} because it describes Alfv\'en perturbations without any restrictions on the wavelength}.

\section*{Acknowledgements}

The study was supported in part by the Ministry of Education and Science of Russia by State assignment to educational and research institutions under Project No. FSSS-2020-0014 and No. 0023-2019-0003, and by RFBR, project number 20-32-90018. CHIANTI is a collaborative project involving George Mason University, the University of Michigan (USA), University of Cambridge (UK) and NASA Goddard Space Flight Center (USA).

\section*{Data Availability}

The data underlying this article will be shared on reasonable request
to the corresponding author.



\bibliographystyle{mnras}
\bibliography{refs} 

\begin{thebibliography}{}
\makeatletter
\relax
\def\mn@urlcharsother{\let\do\@makeother \do\$\do\&\do\#\do\^\do\_\do\%\do\~}
\def\mn@doi{\begingroup\mn@urlcharsother \@ifnextchar [ {\mn@doi@}
  {\mn@doi@[]}}
\def\mn@doi@[#1]#2{\def\@tempa{#1}\ifx\@tempa\@empty \href
  {http://dx.doi.org/#2} {doi:#2}\else \href {http://dx.doi.org/#2} {#1}\fi
  \endgroup}
\def\mn@eprint#1#2{\mn@eprint@#1:#2::\@nil}
\def\mn@eprint@arXiv#1{\href {http://arxiv.org/abs/#1} {{\tt arXiv:#1}}}
\def\mn@eprint@dblp#1{\href {http://dblp.uni-trier.de/rec/bibtex/#1.xml}
  {dblp:#1}}
\def\mn@eprint@#1:#2:#3:#4\@nil{\def\@tempa {#1}\def\@tempb {#2}\def\@tempc
  {#3}\ifx \@tempc \@empty \let \@tempc \@tempb \let \@tempb \@tempa \fi \ifx
  \@tempb \@empty \def\@tempb {arXiv}\fi \@ifundefined
  {mn@eprint@\@tempb}{\@tempb:\@tempc}{\expandafter \expandafter \csname
  mn@eprint@\@tempb\endcsname \expandafter{\@tempc}}}

\bibitem[\protect\citeauthoryear{{Aschwanden} \& {Wang}}{{Aschwanden} \&
  {Wang}}{2020}]{2020ApJ...891...99A}
{Aschwanden} M.~J.,  {Wang} T.,  2020, \mn@doi [\apj]
  {10.3847/1538-4357/ab7120}, \href
  {https://ui.adsabs.harvard.edu/abs/2020ApJ...891...99A} {891, 99}

\bibitem[\protect\citeauthoryear{{Banerjee} et~al.,}{{Banerjee}
  et~al.}{2021}]{2021SSRv..217...76B}
{Banerjee} D.,  et~al., 2021, \mn@doi [\ssr] {10.1007/s11214-021-00849-0},
  \href {https://ui.adsabs.harvard.edu/abs/2021SSRv..217...76B} {217, 76}

\bibitem[\protect\citeauthoryear{Belov, Molevich  \& Zavershinskii}{Belov
  et~al.}{2020}]{Belov2020}
Belov S.~A.,  Molevich N.~E.,   Zavershinskii D.~I.,  2020, \mn@doi [Solar
  Physics] {10.1007/s11207-020-01726-9}, 295

\bibitem[\protect\citeauthoryear{Belov, Molevich  \& Zavershinskii}{Belov
  et~al.}{2021a}]{2021R&QE...63..694B}
Belov S.~A.,  Molevich N.~E.,   Zavershinskii D.~I.,  2021a, \mn@doi
  [Radiophysics and Quantum Electronics] {10.1007/s11141-021-10090-y}, \href
  {https://ui.adsabs.harvard.edu/abs/2021R&QE...63..694B} {63, 694}

\bibitem[\protect\citeauthoryear{Belov, {Vasheghani Farahani}, Molevich  \&
  Zavershinskii}{Belov et~al.}{2021b}]{Belov2021_Alf}
Belov S.,  {Vasheghani Farahani} S.,  Molevich N.,   Zavershinskii D.,  2021b,
  \mn@doi [Solar Physics] {10.1007/s11207-021-01850-0}, 296, 98

\bibitem[\protect\citeauthoryear{Belov, Molevich  \& Zavershinskii}{Belov
  et~al.}{2021c}]{Belov2021}
Belov S.~A.,  Molevich N.~E.,   Zavershinskii D.~I.,  2021c, \mn@doi [Solar
  Physics] {10.1007/s11207-021-01868-4}, 296, 122

\bibitem[\protect\citeauthoryear{Bourdin}{Bourdin}{2017}]{Bourdin2017}
Bourdin P.-A.,  2017, \mn@doi [The Astrophysical Journal]
  {10.3847/2041-8213/aa9988}, 850, L29

\bibitem[\protect\citeauthoryear{{Cohen} \& {Kulsrud}}{{Cohen} \&
  {Kulsrud}}{1974}]{Cohen1974}
{Cohen} R.~H.,  {Kulsrud} R.~M.,  1974, \mn@doi [Physics of Fluids]
  {10.1063/1.1694695}, \href
  {https://ui.adsabs.harvard.edu/abs/1974PhFl...17.2215C} {17, 2215}

\bibitem[\protect\citeauthoryear{De~Moortel \& Hood}{De~Moortel \&
  Hood}{2004}]{DeMoortel2004}
De~Moortel I.,  Hood A.~W.,  2004, \mn@doi [A\&A] {10.1051/0004-6361:20034233},
  415, 705

\bibitem[\protect\citeauthoryear{{Dere, K. P.}, {Landi, E.}, {Mason, H. E.},
  {Monsignori Fossi, B. C.}  \& {Young, P. R.}}{{Dere, K. P.}
  et~al.}{1997}]{Dere1997}
{Dere, K. P.} {Landi, E.} {Mason, H. E.} {Monsignori Fossi, B. C.}  {Young, P.
  R.} 1997, \mn@doi [Astron. Astrophys. Suppl. Ser.] {10.1051/aas:1997368},
  125, 149

\bibitem[\protect\citeauthoryear{{Duckenfield}, {Kolotkov}  \&
  {Nakariakov}}{{Duckenfield} et~al.}{2021}]{2021A&A...646A.155D}
{Duckenfield} T.~J.,  {Kolotkov} D.~Y.,   {Nakariakov} V.~M.,  2021, \mn@doi
  [\aap] {10.1051/0004-6361/202039791}, \href
  {https://ui.adsabs.harvard.edu/abs/2021A&A...646A.155D} {646, A155}

\bibitem[\protect\citeauthoryear{{Field}}{{Field}}{1965}]{Field1965}
{Field} G.~B.,  1965, \mn@doi [Astrophys. J.] {10.1086/148317}, \href
  {https://ui.adsabs.harvard.edu/abs/1965ApJ...142..531F} {142, 531}

\bibitem[\protect\citeauthoryear{Gary}{Gary}{2001}]{Gary2001}
Gary G.~A.,  2001, \mn@doi [Solar Physics] {10.1023/a:1012722021820}, 203, 71

\bibitem[\protect\citeauthoryear{{Hollweg}}{{Hollweg}}{1971}]{Hollweg1971}
{Hollweg} J.~V.,  1971, \mn@doi [\jgr] {10.1029/JA076i022p05155}, \href
  {https://ui.adsabs.harvard.edu/abs/1971JGR....76.5155H} {76, 5155}

\bibitem[\protect\citeauthoryear{{Jess}, {Mathioudakis}, {Erd{\'e}lyi},
  {Crockett}, {Keenan}  \& {Christian}}{{Jess}
  et~al.}{2009}]{2009Sci...323.1582J}
{Jess} D.~B.,  {Mathioudakis} M.,  {Erd{\'e}lyi} R.,  {Crockett} P.~J.,
  {Keenan} F.~P.,   {Christian} D.~J.,  2009, \mn@doi [Science]
  {10.1126/science.1168680}, \href
  {http://ukads.nottingham.ac.uk/abs/2009Sci...323.1582J} {323, 1582}

\bibitem[\protect\citeauthoryear{{Kohutova, P.}, {Verwichte, E.}  \& {Froment,
  C.}}{{Kohutova, P.} et~al.}{2020}]{Kohutova2020}
{Kohutova, P.} {Verwichte, E.}  {Froment, C.} 2020, \mn@doi [A\&A]
  {10.1051/0004-6361/201937144}, 633, L6

\bibitem[\protect\citeauthoryear{{Kolotkov}, {Nakariakov}  \&
  {Zavershinskii}}{{Kolotkov} et~al.}{2019}]{2019A&A...628A.133K}
{Kolotkov} D.~Y.,  {Nakariakov} V.~M.,   {Zavershinskii} D.~I.,  2019, \mn@doi
  [\aap] {10.1051/0004-6361/201936072}, \href
  {https://ui.adsabs.harvard.edu/abs/2019A&A...628A.133K} {628, A133}

\bibitem[\protect\citeauthoryear{Kolotkov, Duckenfield  \& Nakariakov}{Kolotkov
  et~al.}{2020}]{Kolotkov_2020}
Kolotkov D.~Y.,  Duckenfield T.~J.,   Nakariakov V.~M.,  2020, \mn@doi [\aap]
  {10.1051/0004-6361/202039095}, 644, A33

\bibitem[\protect\citeauthoryear{Kolotkov, Zavershinskii  \&
  Nakariakov}{Kolotkov et~al.}{2021}]{Kolotkov2021}
Kolotkov D.~Y.,  Zavershinskii D.~I.,   Nakariakov V.~M.,  2021, \mn@doi
  [Plasma Physics and Controlled Fusion] {10.1088/1361-6587/ac36a5}, 63, 124008

\bibitem[\protect\citeauthoryear{Molevich \& Oraevskii}{Molevich \&
  Oraevskii}{1988}]{Molevich88}
Molevich N.~E.,  Oraevskii A.~N.,  1988, Zh. Eksp. Teor. Fiz, 94, 128

\bibitem[\protect\citeauthoryear{Nakariakov \& Kolotkov}{Nakariakov \&
  Kolotkov}{2020}]{Nakariakov2020}
Nakariakov V.~M.,  Kolotkov D.~Y.,  2020, \mn@doi [Annu Rev of Astron and
  Astrophys] {10.1146/annurev-astro-032320-042940}, 58, 441

\bibitem[\protect\citeauthoryear{Nakariakov, Afanasyev, Kumar  \&
  Moon}{Nakariakov et~al.}{2017}]{Nakariakov_2017}
Nakariakov V.~M.,  Afanasyev A.~N.,  Kumar S.,   Moon Y.-J.,  2017, \mn@doi
  [The Astrophysical Journal] {10.3847/1538-4357/aa8ea3}, 849, 62

\bibitem[\protect\citeauthoryear{{Parker}}{{Parker}}{1953}]{Parker1953}
{Parker} E.~N.,  1953, \mn@doi [Astrophys. J.] {10.1086/145707}, \href
  {https://ui.adsabs.harvard.edu/abs/1953ApJ...117..431P} {117, 431}

\bibitem[\protect\citeauthoryear{{Priest}}{{Priest}}{2014}]{Priest2014}
{Priest} E.,  2014, {Magnetohydrodynamics of the Sun}.
Cambridge University Press, \mn@doi{10.1017/CBO9781139020732}

\bibitem[\protect\citeauthoryear{{Sabri}, {Farahani}, {Ebadi}  \&
  {Poedts}}{{Sabri} et~al.}{2020}]{2020NatSR..1015603S}
{Sabri} S.,  {Farahani} S.~V.,  {Ebadi} H.,   {Poedts} S.,  2020, \mn@doi
  [Scientific Reports] {10.1038/s41598-020-70995-y}, \href
  {https://ui.adsabs.harvard.edu/abs/2020NatSR..1015603S} {10, 15603}

\bibitem[\protect\citeauthoryear{Scalisi, Oxley, Ruderman  \&
  Erd{\'{e}}lyi}{Scalisi et~al.}{2021}]{Scalisi_2021}
Scalisi J.,  Oxley W.,  Ruderman M.~S.,   Erd{\'{e}}lyi R.,  2021, \mn@doi [The
  Astrophysical Journal] {10.3847/1538-4357/abe8db}, 911, 39

\bibitem[\protect\citeauthoryear{Srivastava et~al.,}{Srivastava
  et~al.}{2017}]{Srivastava2017}
Srivastava A.~K.,  et~al., 2017, \mn@doi [Scientific Reports]
  {10.1038/srep43147}, 7

\bibitem[\protect\citeauthoryear{{Vasheghani Farahani} \& {Hejazi}}{{Vasheghani
  Farahani} \& {Hejazi}}{2017}]{2017ApJ...844..148V-2}
{Vasheghani Farahani} S.,  {Hejazi} S.~M.,  2017, \mn@doi [\apj]
  {10.3847/1538-4357/aa7da5}, \href
  {http://ukads.nottingham.ac.uk/abs/2017ApJ...844..148V} {844, 148}

\bibitem[\protect\citeauthoryear{{Vasheghani Farahani, S.}, {Nakariakov, V.
  M.}, {Verwichte, E.}  \& {Van Doorsselaere, T.}}{{Vasheghani Farahani, S.}
  et~al.}{2012}]{2012A&A...544A.127V}
{Vasheghani Farahani, S.} {Nakariakov, V. M.} {Verwichte, E.}  {Van
  Doorsselaere, T.} 2012, \mn@doi [A\&A] {10.1051/0004-6361/201219569}, 544,
  A127

\bibitem[\protect\citeauthoryear{Vasheghani~Farahani, Nakariakov,
  Van~Doorsselaere  \& Verwichte}{Vasheghani~Farahani
  et~al.}{2011}]{2011A&A...526A..80V}
Vasheghani~Farahani S.,  Nakariakov V.~M.,  Van~Doorsselaere T.,   Verwichte
  E.,  2011, \mn@doi [Astron. Astrophys] {10.1051/0004-6361/201016063}, 526,
  A80

\bibitem[\protect\citeauthoryear{{Verwichte}, {Nakariakov}  \&
  {Longbottom}}{{Verwichte} et~al.}{1999}]{Verwichte1999}
{Verwichte} E.,  {Nakariakov} V.~M.,   {Longbottom} A.~W.,  1999, \mn@doi
  [Journal of Plasma Physics] {10.1017/S0022377899007771}, \href
  {https://ui.adsabs.harvard.edu/abs/1999JPlPh..62..219V} {62, 219}

\bibitem[\protect\citeauthoryear{Zanna, Dere, Young  \& Landi}{Zanna
  et~al.}{2020}]{Delzanna2020chianti}
Zanna G.~D.,  Dere K.~P.,  Young P.~R.,   Landi E.,  2020, CHIANTI -- an atomic
  database for emission lines -- Paper XVI: Version 10, further extensions
  (\mn@eprint {arXiv} {2011.05211})

\bibitem[\protect\citeauthoryear{Zavershinskii, Kolotkov, Nakariakov, Molevich
  \& Ryashchikov}{Zavershinskii et~al.}{2019}]{Zavershinskii2019}
Zavershinskii D.~I.,  Kolotkov D.~Y.,  Nakariakov V.~M.,  Molevich N.~E.,
  Ryashchikov D.~S.,  2019, \mn@doi [Phys Plasmas] {10.1063/1.5115224}, 26,
  082113

\bibitem[\protect\citeauthoryear{Zavershinskii, Molevich, Riashchikov  \&
  Belov}{Zavershinskii et~al.}{2020}]{Zavershinskii2020}
Zavershinskii D.~I.,  Molevich N.~E.,  Riashchikov D.~S.,   Belov S.~A.,  2020,
  \mn@doi [Phys. Rev. E] {10.1103/PhysRevE.101.043204}, 101, 043204

\bibitem[\protect\citeauthoryear{Zhugzhda}{Zhugzhda}{1996}]{Zhugzhda96}
Zhugzhda Y.~D.,  1996, \mn@doi [Phys Plasmas] {10.1063/1.871836}, 3, 10

\makeatother
\end{thebibliography}


\bsp	
\label{lastpage}
\end{document}